\documentclass[acmtog]{acmart}
\author{Jerry Jinfeng Guo}
\email{j.guo-3@tudelft.nl}
\author{Elmar Eisemann}
\email{e.eisemann@tudelft.nl}
\affiliation{%
	\department{Computer Graphics and Visualization} % Not in the John Doe Institute!
	\department{Department of Intelligent Systems}
	\department{Faculty of EEMCS}
	\institution{Delft University of Technology}
	\country{the Netherlands}}

\usepackage{booktabs} % For formal tables

\usepackage[ruled,noend]{algorithm2e} % For algorithms

\SetAlFnt{\small}
\SetAlCapFnt{\small}
\SetAlCapNameFnt{\small}
\SetAlCapHSkip{0pt}

\usepackage{enumitem} % for '\newlist' and '\setlist' macros

\setcopyright{rightsretained}
\begin{document}
% Title portion
\title[Geometric Sample Reweighting]{Geometric Sample Reweighting for Monte Carlo Integration}

%Importance Sampling Revisited: \\ geometric Sample Reweighting for Monte Carlo Integration \\ with Applications to Rendering Problems}

% === Custom Commands =========
\renewcommand{\d}[1]{\ensuremath{\mathrm{d\,}\!{#1}}}

\newlist{condenum}{enumerate}{1} % 'condenum': a new, enumerate-like list env.
\setlist[condenum]{label=\bfseries Condition \arabic*.,
	ref=\arabic*, wide}

\newlist{property}{enumerate}{1} % 'condenum': a new, enumerate-like list env.
\setlist[property]{label=\bfseries Property \arabic*.,
	ref=\arabic*, wide}

% === Abstract ================
\begin{abstract}
We present a general sample reweighting scheme and its underlying theory for the integration of an unknown function with low dimensionality. Our method produces better results than standard weighting schemes for common sampling strategies, while avoiding bias. Our main insight is to link the weight derivation to the function reconstruction process during integration. The implementation of our solution is simple and results in an improved convergence behavior. %, which we link to a weight derivation per sample. %Additionally, we present a related sampling strategy. %that improves upon the common practice of sampling according to the function value.
We illustrate its benefit by applying our method to multiple Monte Carlo rendering problems.
%	Our scheme can work with any given set of samples, provided there are no duplicated samples. NOT NEEDED, we can kick duplicates out
%	We believe that by properly reweighting samples, we can further gain back control from randomness in stochastic processes such as Monte Carlo integration.
%	We believe our scheme can also benefit other applications outside rendering.
\end{abstract}

% ============================================================================
% End of 00-abstract.tex 

% === Category and keywords ===

\maketitle

%% ==========================Begin of Main Body=============================
%---------------------------------------------------------------------------
\section{Introduction}
\label{sec:intro}
Monte Carlo (MC) techniques form the foundation of realistic image synthesis for decades \cite{cook1984distributed}. The principle is simple: a function is sampled and the samples are combined to approximate its integral. Standard MC is often referred to as \emph{brute-force}, as its implementation is simple but the variance of the estimation can be high and convergence slow.
One method to usually improve the integral approximation is to reconstruct the underlying function from the samples and much previous work devoted its attention to particular cases (e.g., shadows~\cite{egan2009frequency} or depth of field~\cite{soler2009fourier}). In this work, we revisit the reconstruction process. We derive an easy-to-implement algorithm to compute sample weights that generally improves the approximation when compared to standard weights for general MC integration.
%Much work has been devoted to improved sampling schemes to choose better locations for evaluating a function, such as Importance Sampling~\cite{veach1995MIS}. In this context, it becomes necessary to attribute weights to each sample to avoid a bias.

%Further more, we develop an unbiased solution since the direct application of the consistent approach introduces bias.
% derivation and propose a general scheme that effectively reduces variance for MC integration.
Our observation is that standard sample weights are often less accurate for lower sampling rates because they do not properly reflect the integration domain nor the local sample density. Our weighting scheme considers all samples of a given set and defines weights based on a geometric partitioning of a low-dimensional integration domain. It results in a consistent estimator that outperforms standard weighting schemes. A major contribution of our work is the derivation of an unbiased estimator. It builds upon this partitioning and applies to sets of independent and identically distributed uniform random samples or stratified samples. % or samples generated with certain analytically know distributions.
% \elmar{there is probably the need for being able to integrate}
Specifically, we propose a novel weighting scheme that is easy to implement and builds upon a sound theoretical derivation. It integrates well into existing rendering pipelines, can be parallelized in conjunction with the unbiased estimator, and we demonstrate its benefit over existing schemes via several rendering problems.

%An existing approach to reduce variance is importance sampling, which, instead, influences the sample distribution to reveal details that might otherwise be difficult and time consuming to capture. We will provide a formulation of our approach for this context as well.

%%%-------------------------------------------------------------------------
%%\subsection{Contribution}
%%\label{sec:intro:contribution}
%%We list
%The contributions of this work are as follows:
%\begin{itemize}
%	%\item A novel and complete \emph{effective sample density} theory for MC integration
%	\item A general \emph{reweighting} approach for MC integration
%	\item A \emph{Metropolis variant sampling} technique compatible with MC integration using our reweighting scheme
%	\item An application of our reweighting scheme to rendering
%\end{itemize}

%%-------------------------------------------------------------------------
%\subsection{Organisation of the Paper}
%\label{sec:intro:organisation}
We will first cover prior work and MC integration. We then give the motivation behind our approach (Sec. \ref{sec:formulation}) and present the core of our solution (Sec.~\ref{sec:georeweight}). Numerical performance and applications to rendering are presented in Sec.~\ref{sec:results}.
% =========================================================================
% End of 01-introduction.tex 

%---------------------------------------------------------------------------
\section{Background}
\label{sec:related}
%%%%%%%%%%%%%%%%%%%%%%%%%%%%%%%%%%%%%%%%%%%%%%%%%%%%%%%%%%%%%%%%%%%%%%%%%%%%%%%%%%%%%%%%%%%%%%
% Topics outline:
%	Monte Carlo integration for rendering, Cook
%	Multiple importance sampling, Veach
%	Metropolis sampling and MLT, Veach
%	Multidimensional Adaptive Sampling and Reconstruction for Ray Tracing, Hachisuka
%	Optimal Reweighting for Illumination Integral, Richardo
%	Voronoi cell for anti aliasing, Mitchell
%%%%%%%%%%%%%%%%%%%%%%%%%%%%%%%%%%%%%%%%%%%%%%%%%%%%%%%%%%%%%%%%%%%%%%%%%%%%%%%%%%%%%%%%%%%%%%
%
%In this section we provide prior work related to our research.
%%To make the notoriously naughty bibtex work, we hereby include \cite{Buhmann:1998:DCQ}, \cite{Fellner-Helmberg93}, \cite{FolDamFeiHug.etal93}, \cite{Kobbelt97-USHDR}, \cite{Lafortune97-NARF}, \cite{Lous90} and \cite{Seidel93}.
%%For now, we proudly skip this part as it requires excessive amount of time to write.
%
%\paragraph*{Background}

%IS this needed here?
%
%\paragraph*{Monte Carlo Integration for Rendering}
\paragraph{MC methods}
Since the 80s~\cite{cook1984distributed}, MC integration plays a major role in rendering complex effects, such as motion blur, depth of field, and soft shadows. The complete light transport is described by the \emph{rendering equation}~\cite{kajiya1986rendering}, which can be solved using \emph{path tracing} as an associated MC solution.
Nevertheless, not all samples taken during the evaluation of an integral contribute strongly to the result. One strategy to modify subsequent sample choices is to rely on previous samples, i.e., a Markov process. Metropolis sampling~\cite{veach1997MLT} can handle complex light path configurations by extensively exploring contributing paths once they are discovered.
%enables exploring function values without additional information.%It treats each image region, rather than single pixels, as an integral.
Multidimensional k-d trees~\cite{hachisuka2008multidimensional, Guo2018} can store samples in a global structure, which can then be used as a means to control future sample placement.%Due to the k-D tree, typically up to 5D functions can be handled in practice.

While standard Monte Carlo (MC) methods solve a definite integration $I = \int_{\Omega} f(x)\d x$ of a function $f$ over a finite support $\Omega\subset \mathbb{R}^d$ by using a random sample set ($\{x_i \in \Omega\}$) with the resulting estimator being $\hat{I}_{MC} = \frac{1}{N} \sum_{i=1}^{N} f(x_i)$,
%\begin{equation}
%\label{eq:MCIntegration}
%\hat{I}_{MC} = \frac{1}{N} \sum_{i=1}^{N} f(x_i),	
%\end{equation}
\emph{importance sampling} influences the sampling process via a probability distribution function (pdf) $p:\Omega \rightarrow \mathbb{R}$~\cite{veach1995MIS}. The resulting unbiased estimator is:
\begin{equation}
\label{eq:MCImportance}
\hat{I}_p = \frac{1}{N} \sum_{i=1}^{N} \frac{f(x_i)}{p(x_i)},
\end{equation}
which effectively weighs samples differently.
Importance sampling is interesting when having knowledge about the scene.
For instance, importance sampling the light source works better in scenes with small or point light sources~\cite{dutre2006advanced, debevec2008rendering}.
Sampling according to the BSDF works better with glossy to highly glossy surfaces\cite{shirley1991physically, ward1992measuring, lafortune1997non}.
Multiple importance sampling (MIS) combines different such sampling strategies~\cite{veach1995MIS}.
%Suppose we have $M$ sampling strategies, the unbiased MC estimator with MIS is given as: $\hat{I}_{MIS} = \sum_{i=1}^{M}\frac{1}{N_i}\sum_{j=1}^{N_i}w_i(x_{i,j})\frac{f(x_{i,j})}{p_i(x_{i,j})}$,
% where $N_i$ corresponds to the number of samples for pdfs $p_i(x)$ and weighting heuristics $w_i(x_{i,j})$.
%MIS is a reweighting process and is proved to be unbiased.

%Further, the weighting of the samples  %However, the emphasis of the work is sample placement and it has to work in a few phases.
\paragraph{Reweighting}
Our solution focuses on the weighting of samples interpreted as an improved function reconstruction.
Different weight definitions have been shown to be beneficial for rendering, e.g., derived in Sobolev spaces~\cite{marques2018optimal}. However, these previous solutions target hemispherical illumination integrals and are not generally applicable to other problems. A reweighting scheme was also proposed for addressing firefly artifacts~\cite{zirr2018re} but the solution is biased and  limited to narrow application scenarios.

Other specialized reconstruction techniques exist, including solutions for soft shadows~\cite{egan2009frequency}, defocus blur~\cite{soler2009fourier}, and motion blur~\cite{egan2011frequency}, which lead to significant improvements.
More complex reconstructions for light fields~\cite{lehtinen2011temporal} have proven very successful but are biased (though consistent).

Our method is independent of the application scenario and unbiased.
It handles general functions and links the weights to \emph{Voronoi} cell volumes. The latter has also been studied in the context of anti-aliasing problems~\cite{mitchell1990antialiasing}, for which the $2$ dimensional voronoi cell volumes bounded with a pixel are directly used as sample weights and leads to improved anti-aliasing effects, but the theory has not been further developed for unbiased solutions, nor generalized to other contexts.
Voronoi cell size has been used as weights for Monte Carlo integration in~\cite{vorechovsk2016AVWMC}, where two ways of treating boundaries have been proposed.
In this work, Voronoi cells of given set of samples within a domain are either bounded and clipped by the domain boundary, or extended by periodically adding auxiliary samples that extend the domain.
Both approaches are shown to improve numerical performance of MC integrations.
However, as we show in the Sec.\ref{sec:georeweight}, directly using Voronoi cell size as weight results in an biased estimate.
Our solution takes advantage of Voronoi tessellation and remains unbiased.

\section{Formulation and Problem Statement}
\label{sec:formulation}

Referring to Eq.~\ref{eq:MCImportance}, the estimator of importance sampling is a sum of function value $f(x_i)$ times a weight $\Delta(x_i)$, generally:
\begin{equation}
\label{eq:importance:volumescaling}
	\Delta(x_i) = \frac{1}{N p(x_i)}.
\end{equation}
%When $p$ has a high value, more samples will be used in this area, leading to a more precise sampling/reconstruction of the function. Nevertheless, these samples with a high value of $p$ will have a relatively low weight and contributes less to the integral than the function value might imply.
Similarly, Riemann integration approximates an integral using function values $f(x_i)$ times a weight $w(x_i)$:
\begin{equation}
\label{eq:weightedRSum}
\hat{I} = \sum_{i=1}^{N} w(x_i)f(x_i).
\end{equation} %$w(x_i) = \frac{1}{N} \frac{1}{p(x_i)}$
The Riemann weights stem from a partitioning of the support $\Omega$ into \emph{hypervolumes}.
In 1D, these hypervolumes are intervals.
Each hypervolume contains exactly one sample and its volume defines the sample's weight.
%Each sample lies in exactly one
%Here, $w(x_i)$ have a geometric interpretation; they are \emph{hypervolumes} a set of points in space containing and being associated to a sample $x_i$. In 1D, these hypervolumes are intervals containing sample points and the boundary of $\Omega$. The hypervolumes partition the support $\omega$ and each hypervolume $i$ contains has integral $w(x_i)$ and contains $x_i$. Then we can rewrite the function integral approximation as:
%where $w(x_i)$ is the weight associated with sample $x_i$.
%The weight could be, e.g., $w(x_i)=\frac{1}{N p(x_i)}$ as in Equation \ref{eq:importance:volumescaling}.

\begin{figure}
	\centering
	\includegraphics[width=\linewidth]{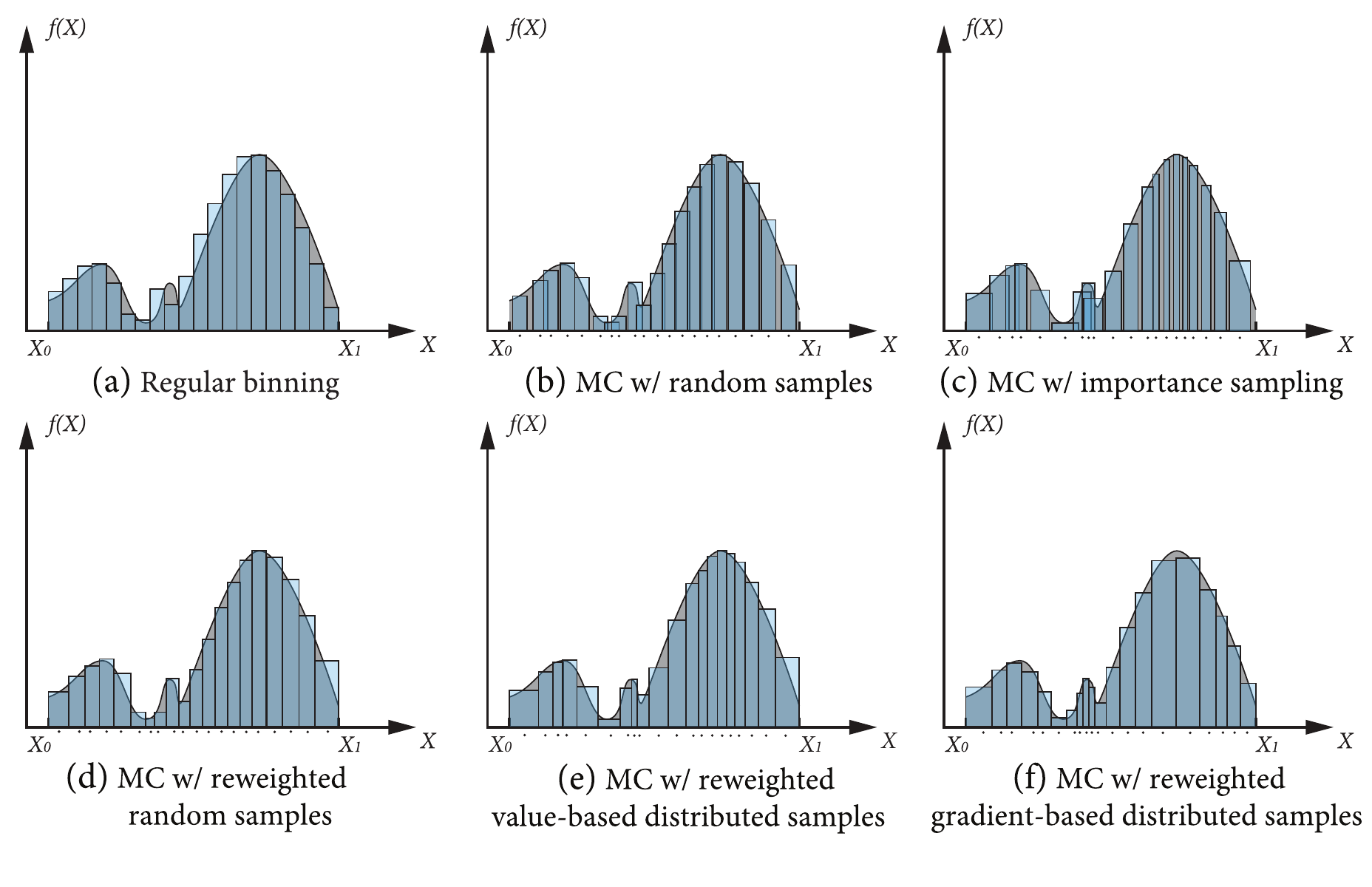}
	\caption{Top row: Three integration methods using the same amount of function evaluations (i.e., $20$ samples): (a) Riemann sum through regular binning (according to right side value)
	% \elmar{is it a problem that typically it would be upper/lower}; 
	(b) MC integration using uniform random samples; (c) MC integration using samples that are distributed according to a pdf w.r.t. function value. Notice that in (a) and (b), the associated bin widths are equal, i.e., $\frac{1}{20}$. Bin widths in (c) are adjusted according to its density determined during sample generation. Notice also the overlaps and gaps between sample bins as illustrated in (b) and (c). Bottom row: illustrations of our methods: (d) uniform random samples with our reweighting; (e) samples distributed according to function value with our reweighting; and (f) samples distributed according to function gradient with our reweighting. Notice the absence of gaps/overlaps and bin widths being adjusted according to sample positions.}
	\label{fig:importance}
\end{figure}

The weights $\Delta(x_i)$ are typically easy to compute but cannot be considered hypervolumes; they would overlap or introduce gaps and cannot easily be linked to a partitioning of $\Omega$. Only with increasing number, due to the stochastic nature of the process, when the samples densely cover the support $\Omega$, the difference in the weight definitions becomes negligible.
%Although quasi MC methods with low discrepancy samples can help improve on this, the noticeable patterns in results are less desirable.
See Fig.~\ref{fig:importance} (a), (b) and (c) for an illustration.
% \elmar{is case c not weird because that could enable importance sampling}
In consequence, especially for low sample counts, the weights do not well reflect an approximation of the function.

%\begin{itemize}
%	\item Ranges in $\Omega$, where density function $p(x)$ has higher values receive more samples;
%	\item Samples with a higher density function value $p(x)$ have a lower hypervolume $\Delta(x)$;
%	\item The sum of all hypervolumes associated with the samples equals the total hypervolume of the support $\Omega$.
%\end{itemize}
%At this point it is worth mentioning that the common practice for importance sampling is to put more samples where the actual function values $f(\cdot)$ are higher.
%This is clearly suboptimal, as high value regions are not necessary regions with more details worth exploring and vice versa.
%Intuitively, one would like more samples where the function varies more and less samples where function remains more or less constant.

\section{Geometric Sample Reweighting}
\label{sec:georeweight}

Our goal is to associate weights to samples that define an improved function reconstruction during the integration.
We will first define a consistent solution, inspired by Riemann integration.
This solution is independent of the sampling pattern and can be applied on any sample set as a post process to improve the approximation.
This reweighting is consistent, but not unbiased for all sampling strategies.
We then propose a modification to obtain an unbiased estimator for the cases of uniform random samples that are \emph{independent and identically distributed} (i.i.d.), and samples with stratification. See Sec~\autoref{sec:conclusion} for the possibility to generalize our method for samples generated with an analytically known pdf.

\subsection{Consistent Estimator}
Riemann integration typically assumes a regular partitioning of the domain. Using a \emph{Voronoi diagram} of the sample points, it is possible to partition the domain $\Omega$ to take sample density into account. A Voronoi diagram is a partition into regions such that the points in each region share the same closest sample location. It can be shown that the Voronoi cell corresponds to the intersection of half spaces defined by hyperplanes that are equidistant to two sample points. The theory of Voronoi diagrams is beyond the scope of this paper but more details can be found in~\cite{aurenhammer1991voronoi, de1997computational}.

In our case of a $D$ dimensional problem setting, the diagram will be bounded by the hypercube $(0,1)^D$, the domain from which samples are drawn. The volume of each Voronoi cell determines the corresponding sample weight and given that the cells are intersections of half-spaces, they are convex and their volume can be easily computed.
The resulting estimator of our approach is $\hat{I}_{CON} = \sum_{i=1}^{N} w_{CON}(x_i)f(x_i)\text{, where } w_{CON}(x_i) = \frac{\left | V_i\right |}{\left | \Omega\right |}$.
% \elmar{I just realize - uniform in sample space is not uniform in path space...}

Implicitly, this construction approximates the integrand via a piecewise-constant representation. Intuitively, to take the most benefit from this interpretation, samples should be chosen with respect to the gradient of the function.
Fig.~\ref{fig:importance} shows an illustration of this strategy. % - numerical results will be presented in Sec.~\ref{sec:results}.
In principle, even more advanced approximations could be used, yet it turns out that such weight definitions, while consistent, lead to a biased estimate. In the following, we will show the reasoning behind this and derive an unbiased estimator for i.i.d.\ uniform sampling and stratified sampling.

\subsection{Deriving an Unbiased Estimator}
\label{sec:unbiasedEstimator}

\subsubsection*{i.i.d. Uniform Samples}
The reason the direct use of the Voronoi cells' volume is biased is due to the samples whose cell shares a boundary with the domain boundary. To illustrate this situation, we will first consider the 1D case with a set $X$ of $N$ (with $N > 3$) i.i.d.\ \textbf{uniform} samples $X := \left\lbrace x_i \in (0,1)\right\rbrace$ before generalizing to $D$ dimensions, and then to stratified sampling.

\paragraph{One Dimension} Let us assume that the one-dimensional sample set $X$ is sorted from smallest to largest value. We are interested in the expected extent of each Voronoi cell, for which we need to derive the expected distance between two adjacent samples. For this reason, we first determine the expected position of sample $x_i$.

From \emph{order statics}~\cite{david2004order}, we know that the distribution of the $i$-th i.i.d.\ sample follows the \emph{beta} distribution, i.e.,
\begin{equation*}
	p_i(x) = \frac{x^{i-1}(1-x)^{N-i} }{\int_{0}^{1}t^{i-1}(1-t)^{N-i}\d t}.
\end{equation*}
The expected position of the ordered $i$-th sample $x_i$ is then:
\begin{equation*}
\begin{split}
	E[ x_i] &= \int_{0}^{1} x \cdot \frac{x^{i-1}(1-x)^{N-i} }{\int_{0}^{1}t^{i-1}(1-t)^{N-i}\d t}\d x \\
			&= N \cdot \binom{N-1}{i-1} \cdot \int_{0}^{1} x \cdot x^{i-1}(1-x)^{N-i} \d x \\
			&= N \cdot \binom{N-1}{i-1} \cdot \int_{0}^{1} x^i \cdot \sum_{k=0}^{N-i}  \binom{N-i}{k} 1^{N-i-k} (-x)^{k} \d x \\
			&= N \cdot \binom{N-1}{i-1} \cdot \int_{0}^{1} \sum_{k=0}^{N-i} \binom{N-i}{k} (-1)^k x ^{k+i} \d x \\
			&= N \cdot \binom{N-1}{i-1} \cdot \sum_{k=0}^{N-i}\binom{N-i}{k}\frac{(-1)^k}{i+k+1} \\
			&= N \cdot \frac{(N-1)!}{(N-i)!(i-1)!} \cdot \frac{\Gamma(i+1)\Gamma(N-i+1)}{\Gamma(N + 2)} \\
			&= N \cdot \frac{(N-1)!}{(N-i)!(i-1)!} \cdot \frac{i!(N-i)!}{(N+1)!} = \frac{i}{N+1}.
\end{split}
\end{equation*}

Consequently, we have $E[|x_i-x_{i+1}|]=\frac{1}{N+1}$ for $i=1$ to $N-1$ and a similar condition holds for $E[x_1-0]=E[1-x_N]=\frac{1}{N+1}$.
The expected weight is then $\frac{1}{N+1}$ for samples $x_i$ with $i=2$ to $N-1$ and $\frac{3}{2}\frac{1}{(N+1)}$ for samples $x_1$ and $x_N$. The latter weights are larger due to the intervals containing the two boundaries of the domain. Using these weights directly, leads to a consistent but biased estimator.

%To attribute the same weight to each sample, we need to perform a per-sample correction, such that the expected contribution of any sample $x_i$ is $\frac{1}{N} f(x_i)$.
To render the estimator unbiased, we introduce a per-sample correction coefficient $C$:
\begin{equation}
\label{eq:MCGeoRW}
	\hat{I}_{GR} = \sum_{i=1}^{N} w_{GR}(x_i)f(x_i)\text{, where } w_{GR}(x_i) = \frac{\left | V_i\right |}{C(x_i)\left | \Omega\right |}.
\end{equation}
These factors have to be carefully chosen --- for instance, $C=1$ would lead to the previously-derived consistent but biased result.
The correction coefficient should indeed modify the expected contribution of a sample $x_i$ to equal $\frac{1}{N}f(x_i)$.
Following the weight derivation, an unbiased estimator in 1D, we would then define $C(x_1)=C(x_N)=\frac{2(N+1)}{3 N}$ and $C(x_i)=\frac{N+1}{N}$ for all other samples.
As most samples still share an identical correction factor, it keeps us close to the interpretation of the Voronoi cell volume.
In higher dimensions, the definition is less straightforward.

\paragraph{D Dimensions} To derive the correction coefficient $C$ from Eq.\ref{eq:MCGeoRW} in $D$ dimensions, we assume a set of $N$ (with $N\geq3^D$, i.e., intuitively, this results in at least one inner point and two boundary points along each dimension) samples in $\Omega = \left(0, 1\right)^D$. We define the \emph{boundary order} $b(x_i)$ of a sample as the amount of cell boundaries of its Voronoi cell that are part of the domain boundary. For instance, in the above one dimensional example, $b(x_1)=b(x_N)=1$, and for all other sample points, we have $b(x_i)=0$.

The cardinality of samples of order $d$ is defined as: $|X_d| = \\\textup{card} \left\{ b(x_i)=d, \forall i \in \left[1, N\right]\right\}$.
For such a sample set of $N$ samples, the expected cardinality of samples of order $d$ is $E[ |X_d| ] = \binom{D}{d} (\sqrt[D]{N}-2)^d 2^{D-d}$. This formula is the $d$-th term in the bionomial expansion of $[$($\sqrt[D]{N}-2$)$ + 2]^D$. To understand this result, one should recall that the expected position of all samples forms a regular grid. Thus, this grid will have a resolution of $n = \sqrt[D]{N}$ along each axis. Starting with one axis, we would find $n$ samples with two boundary samples of order one and all others samples are inner points of order zero. Repeating these samples $n$ times along a new dimension will increment the order of the first repeated set of samples and the last, as these represent a new boundary along this dimension. For all other samples, their boundary order remains unchanged. This process can be done for all $D$ dimensions, thus implying the binomial expansion.

To achieve an unbiased estimator, we first compute the expected Voronoi volume $E\left[V_i\right]$ for a sample $x_i$.
For $D$ dimensions, we have $D+1$ boundary orders from $0$ to $D$.
As we are dealing with an i.i.d.\ uniform distribution,
in each dimension, we have $n-1$ intervals between samples and two intervals with the boundary, leading to a total of $n + 1$ intervals.
Therefore, we have:
\begin{align*}
	E\left[V_i\right] &= \left(\frac{3}{2}\right)^{b(x_i)}\frac{\left|\Omega\right|}{n+1}
\end{align*}

Again, for unbiasedness, we need $E\left[w_{GR}(x_i)\right]=1/N$, thus each sample should expectedly contribute equally.
The following definition of the correction coefficients fulfills this property:
\begin{equation}
	C(x_i) = \left(\frac{3}{2}\right)^{b(x_i)} \frac{N}{n+1},
\end{equation}
because
%\begin{align*}
%	\int p_{\textup{eff}} &= \frac{1}{N} E\left[\frac{\left|\Omega\right|C(x_i^d)}{N\left|V_i\right|} \right] = \frac{\left|\Omega\right|}{N} \sum |x_i^d| \frac{C(x_i^d)}{N\times E\left[\left|V_i\right|\right]} \\
%	&= \frac{\left|\Omega\right|}{N} \sum \binom{D}{d} (n-2)^d 2^{D-d} \left(\frac{3}{2}\right)^d \frac{N}{n+1} \frac{1}{N\times E\left[V_i\right]} \\
%	&= \frac{\left|\Omega\right|}{N} N = \left|\Omega\right| = 1.
%\end{align*}
%Therefore, we have:
\begin{align*}
	E\left[w_{GR}(x_i)\right] &= E\left[ \frac{\left|V_i\right|}{C(x_i)\left| \Omega \right|} \right]= \frac{1}{\left|\Omega\right|} E\left[ \frac{\left|V_i\right|}{C(x_i)} \right] \\
	&= \frac{1}{\left|\Omega\right|} \frac{\left(\frac{3}{2}\right)^{b(x_i)}\frac{\left|\Omega\right|}{n+1}}{\left(\frac{3}{2}\right)^{b(x_i)}\frac{N}{n+1}} = \frac{1}{\left|\Omega\right|} \times \frac{\left|\Omega\right|}{N} = \frac{1}{N}.	  \hfill\ensuremath{\blacksquare}
\end{align*}
% Hence, $E\left[w_{GR}(x_i)\right]=1/N$. \hfill\ensuremath{\blacksquare}
%Therefore $E\left[ \hat{I}_{GR} \right] = I$.

\subsubsection*{Stratified Samples}
\label{sec:unbiasedEstimator:stratification}
The extension to stratified sampling is relatively straightforward, as each stratum is considered an independent unit. This means that the function is independently integrated in each stratum and its whole range is a composition of these units. In consequence, the boundary observation now applies to the boundary of each stratum. For a sample set $X$ of size $N$ generated with $S$ strata, each stratum is expected to contain $\frac{N}{S}$ samples.
Let $n = \sqrt[D]{N}$ and $s = \sqrt[D]{S}$, then the integration problem for a stratum with $\frac{N}{S}$ samples would imply the correction coefficients to be:
\begin{equation}
C(x_i) = \left(\frac{3}{2}\right)^{b(x_i)} \frac{N}{n+s}.
\end{equation}

\section{Results}
\label{sec:results}
In this section, we first show the numerical performance (Sec.~\ref{sec:results:numerical}) of our scheme and show its application to a few rendering scenarios (Sec.~\ref{sec:results:application}). In all tests, we compare four different estimators:
\begin{enumerate}
	\item Standard MC (i.i.d.\ uniform sampling)
	\item Our weighted standard MC (i.i.d.\ uniform sampling)
	\item Stratified MC (stratified sampling)
	\item Our weighted stratified MC (stratified sampling)
\end{enumerate}

\subsection{Numerical Performance}
\label{sec:results:numerical}
\begin{figure*}[tbp]
	\centering
	\includegraphics[width=\linewidth]{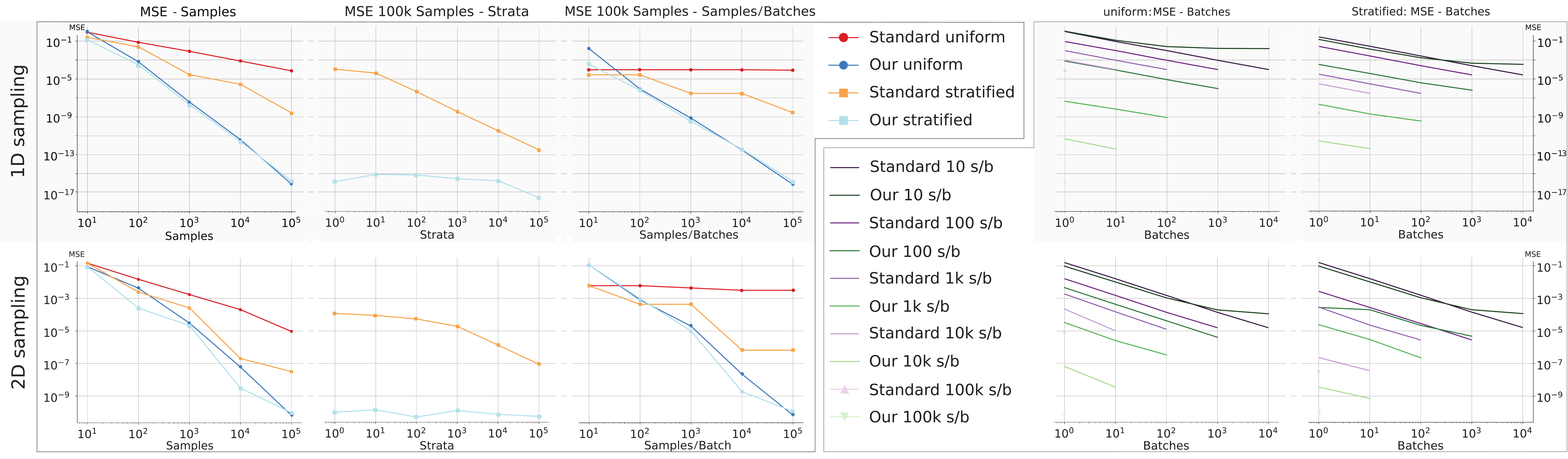}
	\caption{We apply our geometric sample reweighting to one and two dimensional MC integration problems.}
	\label{fig:12DMSE}
\end{figure*}

The numerical performance is tested with two examples: one for a $1$D MC integration and the other for a $2$D MC integration, which are plotted in \autoref{fig:12DMSE}. % and \autoref{fig:2DMSE}.
%\elmar{ADD FUNCTION!}
The $1$D function is given as:
\begin{equation*}
	f(x) = 10\times\begin{cases} 
		\sqrt{-x^2+0.5x} & 0<x<0.25\\
		-\sqrt{-x^2+x-0.1875}+0.25 & 0.25 < x < 0.5\\
		20\times (x-0.5)& 0.5 < x < 0.55\\
		1.0& 0.55 < x < 0.65\\
		-20\times(x-0.7)& 0.65 < x < 0.7\\
		0.1\times \sin (10\pi\cdot(x-0.7))& 0.7 < x < 0.8\\
		0.25\times \sin (10\pi\cdot(x-0.8))& 0.8 < x < 0.9\\
		0.5\times \sin (10\pi\cdot(x-0.9))& 0.9 < x < 1.0
		\end{cases}
\end{equation*}
For the $2$D function, we take the \emph{Lena} image~\cite{munson1996note}.
The functions were chosen to include discontinuities, large-scale variations and small scale changes and led to a representative behavior of several tests that we have performed. 
Generally, the MSE drops as more samples are added (Column $1$). Our solutions outperform standard uniform sampling and even stratified sampling by several orders of magnitude and converges around 1000, 100 times faster respectively in 1D and 100, 10 times faster respectively in 2D.

For the case of stratified sampling, we illustrate different amounts of strata for the same sampling count (Column $2$). Our weighting scheme makes this parameter less important, as it achieves a better function approximation.

We next investigate the impact of distributing samples into batches for which we estimate the function integration separately, before deriving the overall estimate by averaging, which would typically be the case for distributed computations.
First, we fixate the amount of samples to $100$K (Column $3$). Notice that the performance of standard uniform sampling remains invariant with respect to the amount of samples per batch, as it is already an averaging process. Our solution results in a better approximation for more samples per batch, as it will approximate the function more faithfully, as expected. Similarly, stratified sampling also benefits from more samples per batch, but shows slower convergence.

We also investigate the effect of using different batch sizes for uniform (Column $4$) and stratified sampling(Column $5$). More batches thus means a higher overall sample count and all methods improve with the addition of batches. In all cases, the graphs stop after reaching $100$K samples. Our solution performs best and the graphs also illustrate the convergence over several batches, due to its unbiasedness.

% \begin{figure*}[tbp]
% 	\centering
% 	\includegraphics[width=\linewidth]{img/numerical-analysis-2D}
% 	\caption{We apply our geometric sample reweighting to a two dimensional MC integration problem.}
% 	\label{fig:2DMSE}
% \end{figure*}

%-------------------------------------------------------------------------
\subsection{Application to Rendering}
\label{sec:results:application}

We implemented our method in Mitsuba \cite{Mitsuba}, targeting one and two dimensional integration problems, namely motion blur (Sec.~\ref{sec:results:application:temporal}), dispersion (Sec.~\ref{sec:results:application:spectral}), depth of field (Sec.~\ref{sec:results:application:lens}) and illumination integrals (Sec.~\ref{sec:results:application:illumination}).
We evaluate MSE and visual appearance, as well as convergence behavior.
For all implementations, our reweighting operates at a per-pixel level.
We apply our method on the level of primary samples, thus all applicable local importance sampling techniques are utilized throughout the pipeline.
%For spectral sampling, importance sampling did not improve the results and was omitted.

\subsubsection{Motion Blur}
\label{sec:results:application:temporal}
\begin{figure*}[tbp]
	\centering
	\mbox{} \hfill
	\includegraphics[width=\textwidth]{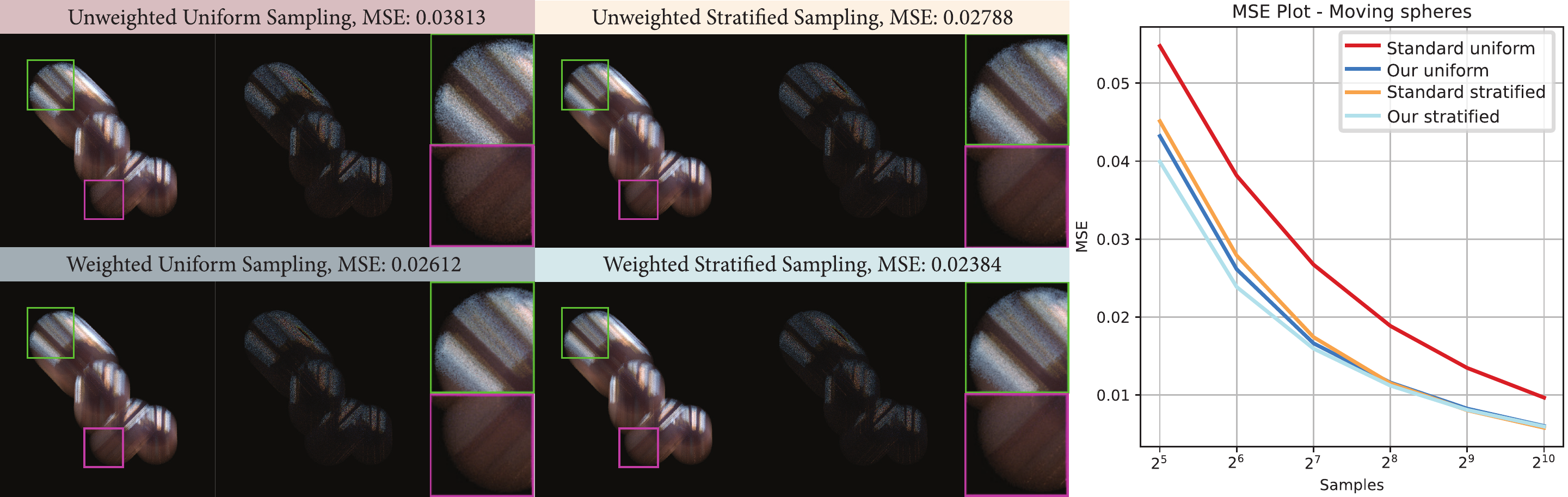}
	\caption{Four highly glossy spheres moving in different directions with $64$ samples per pixel. In each subfigure: corresponding render, difference with reference and highlighted regions.}
	\label{fig:result:sphere}
\end{figure*}

\begin{figure*}[tbp]
	\centering
	\mbox{} \hfill
	\includegraphics[width=\textwidth]{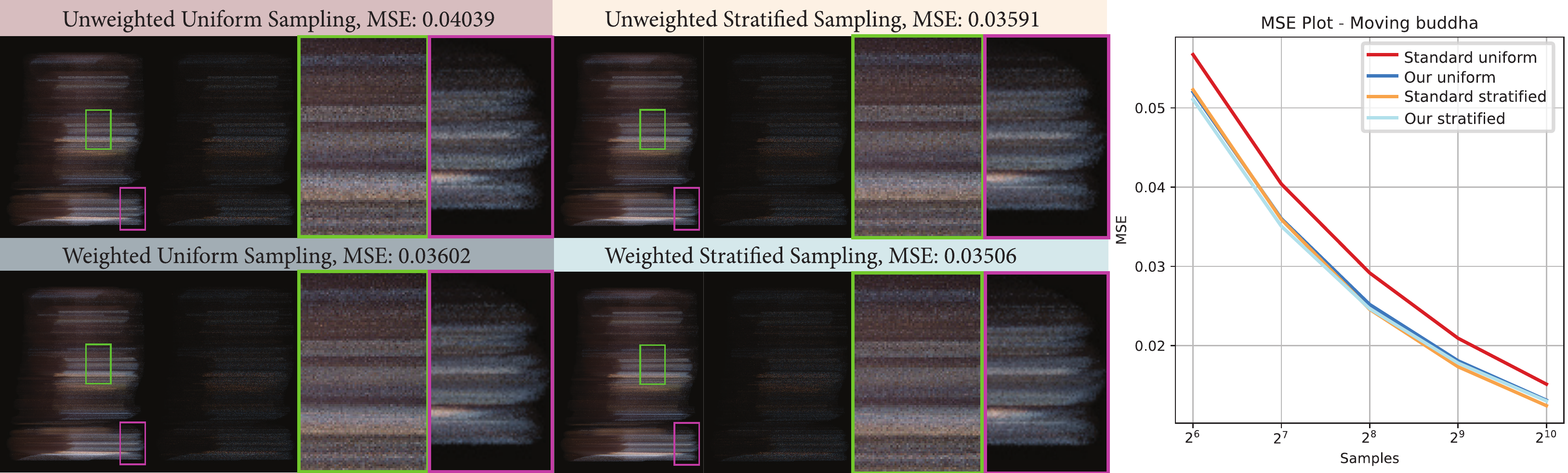}
	\caption{Highly glossy Buddha moving horizontally with $128$ samples per pixel. In each subfigure: corresponding render, difference with reference and highlighted regions.}
	\label{fig:result:buddha}
\end{figure*}

To simulate motion blur, distribution rendering samples the time domain:
For a pixel $(i,j)$, the luminance $L_{(i,j)}$ is given by:
\begin{equation*}
	L_{(i,j)} = \int_{t_{\textup{open}}}^{t_{\textup{close}}}f_{(i,j)}(t) \d t,
\end{equation*}
with $t_{\textup{open}}$ and $t_{\textup{close}}$ being the shutter opening and closing time and $f$ incorporating the shutter function.
%The associated estimator for actual computation is thus given as:
%\begin{equation}
%\label{eq:temporal}
%	\hat{L}_{(i,j)} = \frac{1}{N} \sum_{k=1}^{N} f_{(i,j)}(t_k).
%\end{equation}
%Using Equation \ref{eq:MCGeoRW}, we add our geometric sample reweighting to Equation \ref{eq:temporal} and get:
%\begin{equation}
%\hat{L}_{(i,j),GR} = \sum_{k=1}^{N} \frac{\left | V(t_k) \right |}{C(t_k)\left | t_{\textup{close}} - t_{\textup{open}}\right |} f_{(i,j)}(t_k).
%\end{equation}
Since time is $1$ dimensional, building a Voronoi partition means sorting and measuring the distance between samples.
We tested our implementation in two scenes with animation (Fig. \ref{fig:result:sphere} and \ref{fig:result:buddha}).

%-------------------------------------------------------------------------
\subsubsection{Spectral Rendering}
\label{sec:results:application:spectral}
Light dispersion can happen at reflective or refractive dielectric materials, leading to effects such as rainbows, resulting from different wavelengths travelling in different directions.
Spectral sampling simulates multiple wavelengths in order to capture such effects.
To reduce the complexity of the additional spectral dimension\cite{bergner2009tool}, \emph{hero wavelength spectral sampling}\cite{wilkie2014hero} can be used as an approximation:
\begin{equation*}
\hat{I}_{(i, j)} = \frac{1}{N}\sum_{k=1}^{N}\frac{f_{(i,j)}(\lambda_k)}{p_{(i,j)}(\lambda_k)}
\end{equation*}

%The MC estimator with geometric reweighting is given as:
%\begin{equation}
%\hat{I}_{i,j} = \sum_{k=1}^{N}\frac{\left | V(\lambda_k) \right |}{C(\lambda_k)\left | \lambda_{max} - \lambda_{min}\right |} f_{(i,j)}(\lambda_k).
%\end{equation}

\begin{figure*}[tbp]
	\centering
	\mbox{} \hfill
	\includegraphics[width=\textwidth]{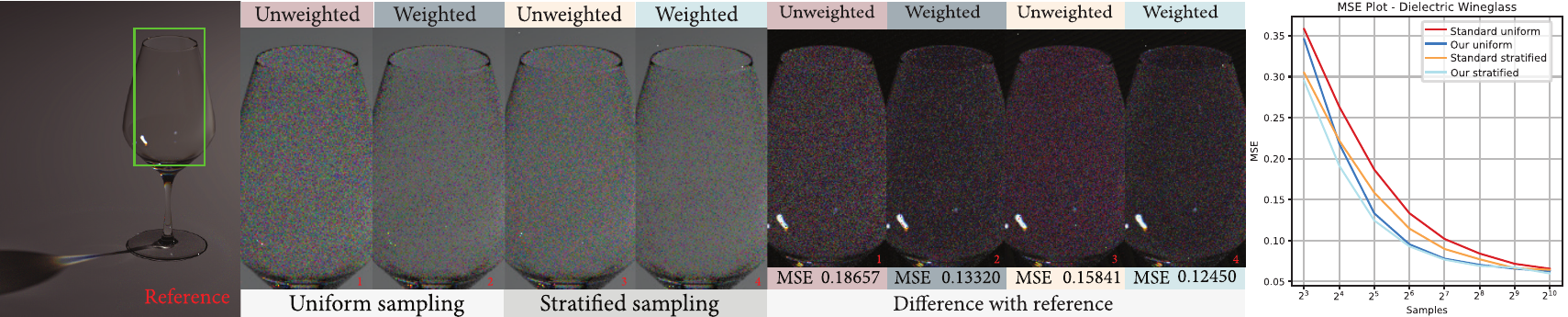}
	\caption{Wineglass with dispersive dielectric materials with $64$ and $32$ samples per pixel. MSE for highlighted four plots are: $0.18657$, $0.13320$, $0.15841$ and $0.12450$ respectively.}
	\label{fig:result:wineglass}
\end{figure*}

\begin{figure*}[tbp]
	\centering
	\mbox{} \hfill
	\includegraphics[width=\textwidth]{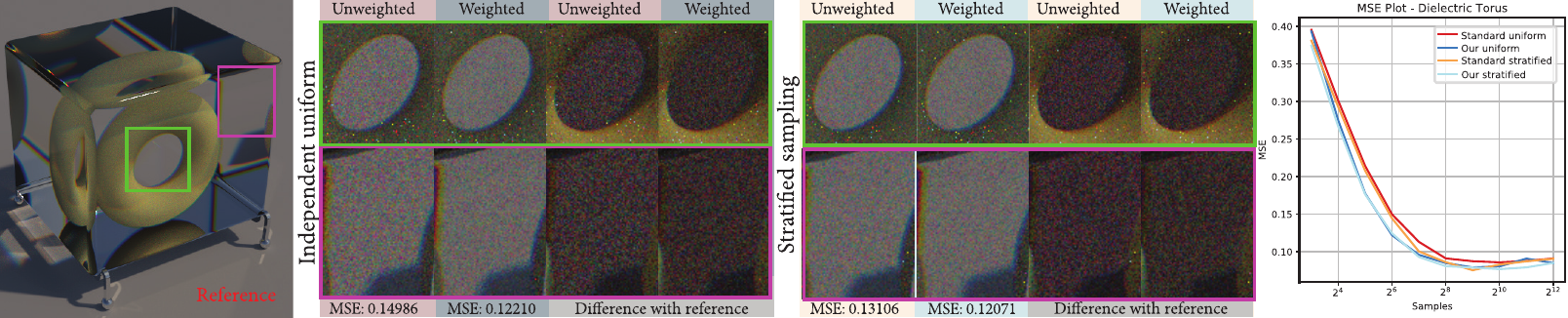}
	\caption{Torus with dielectric materials with $32$ samples per pixel.}
	\label{fig:result:torus}
\end{figure*}

Our implementation of spectral sampling uses $15$-bin wavelengths. Hero wavelength sampling is used with $3$ shifted additional wavelength samples~\cite{wilkie2014hero}.
We tested our method with two scenes configured with dispersive di-electric materials (Fig.~\ref{fig:result:wineglass} and \ref{fig:result:torus}).

As shown in the results, our method brings down colour noise significantly and dispersive regions look much smoother at low sample rate.

%-------------------------------------------------------------------------
\subsubsection{Defocus Blur}
\label{sec:results:application:lens}
A camera with aperture leads to defocus blur/depth of field effects.
The aperture is usually modelled as a $2$D shape, e.g., a square, a circle, or a star, which is sampled to determine the origin of each primary sample ray, which passes through the position on the focal plane corresponding to the current pixel.
For lens aperture $\mathbb{A} \subset \mathbb{R}^2$, we obtain:
\begin{equation*}
	L_{(i,j)} = \int_{\mathbb{A}}f_{(i,j)}(s)\d s.
\end{equation*}

\begin{figure*}[t]
	\centering
	\mbox{} \hfill
	\includegraphics[width=\textwidth]{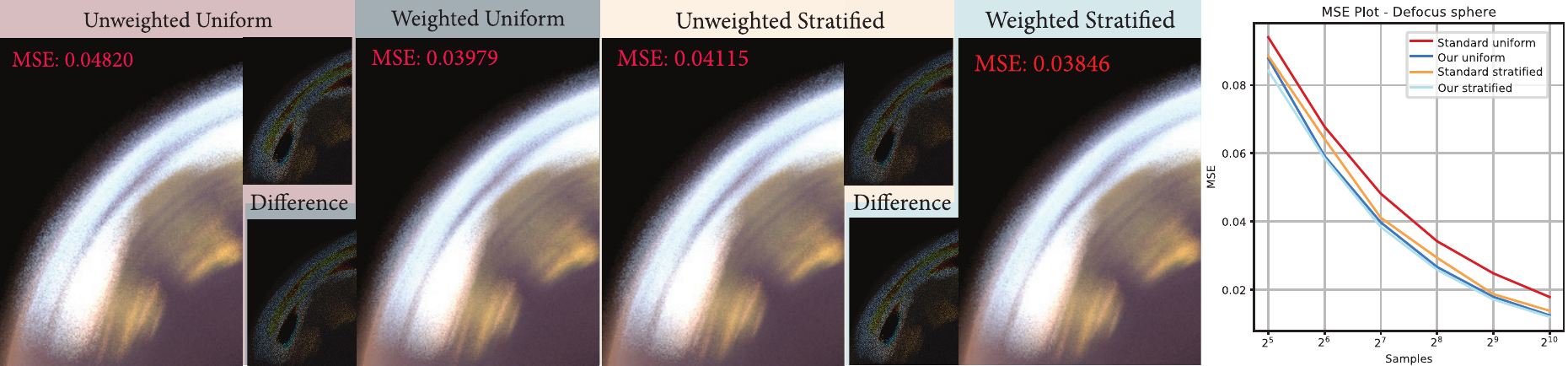}
	\caption{Glossy sphere crossing the focal plane of the camera with $128$ samples per pixel.}
	\label{fig:result:dragon}
\end{figure*}

To determine our weights, we use a $2$D Voronoi diagram based on the aperture samples.
We tested a simple glossy sphere illuminated using an environment map (Fig.~\ref{fig:result:dragon}).

% Similar to the previous experiments, a noticeable difference occurs in the falloff regions and function gradient sampling with our reweighting consistently outperforms the competitors.

%The MC estimator and MC estimator with geometric reweighting are respectively:
%\begin{align}
%	\hat{L}_{(i,j)} 	&= \frac{1}{N} \sum_{k=1}^{N} f_{(i,j)}(s_k), \nonumber \\
%	\hat{L}_{(i,j),GR} 	&= \sum_{k=1}^{N} \frac{\left | V(s_k) \right |}{C(s_k)\left | \mathbb{A} \right |} f_{(i,j)}(s_k).
%\end{align}
%
%%------------------------------------------------------------
%\subsection{Application to Pixel Sampling} % Anti-aliasing}
%\label{sec:application:antialising}
%Pixel Sampling is the technique for anti-aliasing purposes.
%Anti-aliasing is the process of getting rid of alias in pixels, which is usually more severe when pixels are at geometry or texture boundaries.
%The integrand we are seeking to solve for anti-aliasing purposes is the area of a pixel, $\mathbb{P} \subset \mathbb{R}^2$.
%This is also the original application presented in \cite{mitchell1990antialiasing}.
%The pixel luminance for anti-aliasing is given by:
%\begin{equation*}
%	L_{(i,j)} = \int_{\mathbb{P}} f_{(i,j)}(x) \d x.
%\end{equation*}
%
%The MC estimator and MC estimator with geometric reweighting are respectively:
%\begin{align}
%	\hat{L}_{(i,j)} 	&= \frac{1}{N} \sum_{k=1}^{N} f_{(i,j)}(x_k), \nonumber \\
%	\hat{L}_{(i,j),GR} 	&= \sum_{k=1}^{N} \frac{\left | V(x_k) \right |}{C(x_k)\left | \mathbb{P} \right |} f_{(i,j)}(x_k).
%\end{align}

%-------------------------------------------------------------------------
\subsubsection{Direct Illumination}
\label{sec:results:application:illumination}
\begin{figure*}[t]
	\centering
	\mbox{} \hfill
	\includegraphics[width=\textwidth]{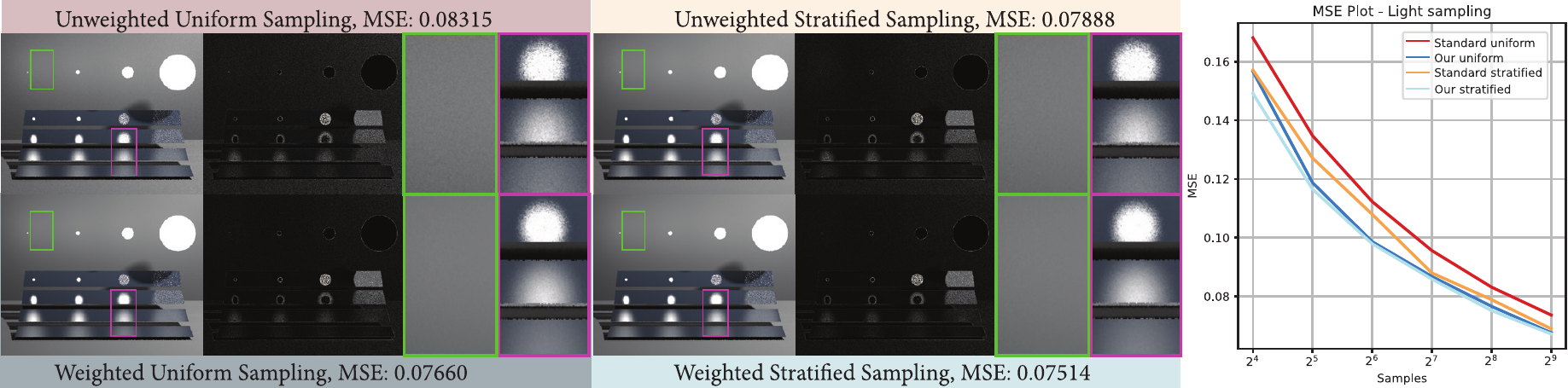}
	\caption{Each image uses $4$ primary rays per pixel and at each scattering event $256$ light samples. Our solution is applied to the $256$ samples.}
	\label{fig:result:lightSampling}
\end{figure*}

%The illumination integral models scattering and is recursive.
% and require multiple bounces/dimensions before it can be considered \emph{solved}.
Leaving out irrelevant terms, the luminance $L_{\textbf{x}}$ at scattering point $\textbf{x}$ with one bounce is given by:
\begin{align*}
	L_{\textbf{x}} 	&= L_e(\textbf{x}) + L_{\textup{direct}} + L_{\textup{indirect}} \\
					&= L_e(\textbf{x}) + \int_{\mathbb{L}} f_s(\textbf{x}) L_e(l\rightarrow \textup{x})  \d l + \int_{\Omega} f_s(\textbf{x}, \omega) L_i(\omega) \d\omega ,
\end{align*}
where $L_e$ denotes light emission and $l\in\mathbb{L}$ denotes all light sources.
In this application, we use light sampling instead of random rays to ensure that the light source is always sampled.
% In consequence, in this case, uniform sampling performs better than the explorative Metropolis variants.
% We do still observe that the gradient approximation performs better than the value-driven distribution.
Our unbiased reweighting achieves the best convergence and, as shown in the insets, also the smoothest results (Fig.~\ref{fig:result:lightSampling}).

\subsubsection*{Observation}
In all cases, our solution leads to smoother visual result and less black holes in the falloff regions.
From the MSE plots, we can see that standard MC with uniform sampling has the worst performance, while our weighted stratified sampling generally has the best one.
Our method improves both uniform sampling and stratified sampling.
We can also see that even with uniform sampling as input, our weighted uniform sampling not only improves over the unweighted version, but also has a performance that is as good as our weighted stratified sampling.
Precompute sample weights enables a negligible computation overhead.

\section{Conclusion}
\label{sec:conclusion}

The reweighting scheme in this paper enables a better approximation than standard MC weights.
Our solution is general and does not require any prior knowledge about the integrating function.
Implicitly, our method approximates this function via a reconstruction from the samples, but does not introduce a bias in the resulting estimator. We showed its practical benefit for various rendering problems.

While we focus on primary samples that are either i.i.d.\ uniform or stratified in this work, our method can also handle non-uniform sample sets following a distribution of $p(x)$. 
The expected position of the $i$-th sample $x_i$ is then $N \cdot \binom{N-1}{i-1} \cdot \int_{0}^{1} P^{-1}(x) \cdot x^{i-1}(1-x)^{N-i} \d x$, where $P(u) = \int p(u) \d u $. 
Unfortunately, it is necessary to integrate the distribution function. Approximate schemes remain an area of future work. Similarly, using the method in higher dimensions requires the computation of cell volumes in high-dimensional Voronoi diagrames, which can be costly. One could precompute these weights but we left such accelerations as future work.
Finally, it is an exciting opportunity to exploit the generality of our solution to improve other integration problems.

\bibliographystyle{ACM-Reference-Format}
\bibliography{sample-bibliography}

%---------------------------------------------------------------------------
% Appendix
%\appendix
%\input{appendix}

%% ===========================End of Main Body============================== 

\end{document}